\documentclass[sigconf,authorversion,nonacm]{acmart}

\usepackage{algorithm}
\usepackage{algorithmic}
\usepackage{subcaption}
\usepackage{soul}
 \usepackage{multirow}

\newfloat{prompt}{thp}{lop}
\floatname{prompt}{Prompt}

\begin{document}

\title{Social Media and Academia: How Gender Influences Online Scholarly Discourse}

\author{Rrubaa Panchendrarajan}
\affiliation{%
  \institution{Queen Mary University of London}
  \country{United Kingdom}
}

\author{Harsh Saxena}
\affiliation{%
  \institution{Birla Institute of Technology And Science Pilani}
  \country{India}}

\author{Akrati Saxena}
\affiliation{%
  \institution{Leiden Institute of Advanced Computer Science,}
  \city{Leiden University}
  \country{The Netherlands}
}
\email{a.saxena@liacs.leidenuniv.nl}


\begin{abstract}
This study investigates gender-based differences in online communication patterns of academics, focusing on how male and female academics represent themselves and how users interact with them on the social media platform X (formerly Twitter). We collect historical Twitter data of academics in computer science at the top 20 USA universities and analyze their tweets, retweets, and replies to uncover systematic patterns such as discussed topics, engagement disparities, and the prevalence of negative language or harassment. The findings indicate that while both genders discuss similar topics, men tend to post more tweets about AI innovation, current USA society, machine learning, and personal perspectives, whereas women post slightly more on engaging AI events and workshops. Women express stronger positive and negative sentiments about various events compared to men. However, the average emotional expression remains consistent across genders, with certain emotions being more strongly associated with specific topics. Writing-style analysis reveals that female academics show more empathy and are more likely to discuss personal problems and experiences, with no notable differences in other factors, such as self-praise, politeness, and stereotypical comments. Analyzing audience responses indicates that female academics are more frequently subjected to severe toxic and threatening replies. Our findings highlight the impact of gender in shaping the online communication of academics and emphasize the need for a more inclusive environment for scholarly engagement.
\end{abstract}


\keywords{Gender Bias, Scholarly Discourse Analysis, Gender Influence}

\maketitle

\section{Introduction}

Social media platforms, specifically X (previously known as Twitter), have become critical spaces for public discourse, shaping opinions and amplifying voices across the globe. However, they also mirror and potentially exacerbate societal biases, including gender-based inequalities \cite{usher2018twitter, macedo2024gender, alba2022bias, salvador2024unveiling, messias2017white}. Gender biases in communication manifest in various forms, such as differences in engagement patterns, language use, sentiment, and visibility of male and female users. Previous studies have examined gender-related differences in both offline and online communication spaces, including platforms such as Twitter \cite{holmberg2015gender, Garcia_Weber_Garimella_2014}, Stack Overflow \cite{may2019gender}, Wikipedia \cite{man_wikipedia}, and YouTube \cite{youtube_gender}. 

Messias et al. \cite{messias2017white} found that white men are more likely to be followed on Twitter and dominate higher positions in rankings, with women only surpassing men in representation after the top 14\% of most-followed users. This trend extends across specific domains, such as sports, politics, technology, and art, where women tend to have less influence and acquire lower ranks among top influencers~\cite{nilizadeh2016twitter,manzano2019women}. Certain fields—particularly sports, STEM (Science, Technology, Engineering, and Mathematics), and politics—are often socially perceived as male-dominated spaces~\cite{man_wikipedia, hu2021gendered}, which contributes to higher male participation in these discussions. Consequently, in these contexts, women can feel more hesitant to talk about such topics, and it can also make them inclined to certain professions due to gender norms and constructs. 

Skewed narratives and perspectives get reinforced as a consequence and get embedded as a subconscious behavior. To move against this reinforcement, we need to be aware of gender differences to promote more diversity, inclusivity, and fairness in online platforms. The dominance of male narratives on platforms like Twitter/X is not merely a reflection of individual preferences but a product of structural imbalances and gender biases in digital participation~\cite{Garcia_Weber_Garimella_2014}. These biases not only perpetuate stereotypes and marginalize underrepresented voices but also contribute to the creation of inequitable and exclusionary digital spaces. As prior research has shown, majority groups tend to dominate online communication, which often leads to minority users feeling less empowered to express themselves freely~\cite{minority_analysis}. 

The study of gender biases in science and scientific communication has long been a focal point for researchers. In this work, we aim to examine the communication patterns of male and female academicians, including assistant professors, associate professors, and full professors, in computer science from the top 20 universities in the United States. Specifically, we investigate whether male and female academicians represent themselves differently on social media and whether they experience varying interactions with social media users. The primary reason for restricting the scope to the top 20 universities, computer science domains, and specific positions is to ensure the comparability of profiles. 
We collect historical posts of these academicians to understand whether there are any significant differences in the way these genders represent themselves to the audience. We also collect the replies of all their tweets posted in a specific duration to observe if there is any significant difference in the way users respond differently to different genders. We analyze the posts and replies to understand the differences in the topics discussed, emotions and sentiments expressed for different scenarios, and as well as the response of users to these academics over time. 
This research seeks to uncover systemic patterns, such as differences in topics discussed by male and female academicians, disparities in engagement levels (e.g., likes, retweets, replies) for tweets by women versus men, and whether women are disproportionately subjected to negative language or harassment. Identifying and understanding these biases is critical for enhancing the inclusivity and safe space of online platforms and guiding the design of algorithms that determine content visibility and moderation.

The paper is structured as follows. In the next section, we discuss related work highlighting gender biases in social media. In Section 3, we discuss the data collection process and the collected dataset. It is followed by the data analysis and our findings. The paper concludes with the Discussion and Conclusion section that summarizes our insights, limitations, and future directions.

\section{Related Work}

This section reviews research on academic discourse on Twitter and gender biases on Social media platforms, highlighting the differences in usage behavior such as discussed topics and linguistic styles, expression of emotions and sentiments, and the users' response to different genders.

\subsection{Academic Discourse on Social Media}

Regular surveys of U.S. university faculty indicate a steady rise in the professional use of social media platforms \cite{seaman2013social}, as it might lead to better visibility and recognizability by sharing their work and collaboration opportunities. However, academic influence on Twitter is judged through complex social logics that go beyond traditional metrics like the h-index. Recognizability, shared interests, and perceived capacity for meaningful contribution are often more important than formal credentials, or follower counts \cite{stewart2015open}. Researchers use social media platforms to share their publications, and it is important to study their impact on the citation of articles. Klar et al. \cite{klar2020using} observed that tweets about academic articles are positively correlated with citation rates, but there was no significant evidence of gender differences in how research is disseminated via Twitter. Several other studies, including \cite{hayon2019twitter, ozkent2022correlation, demir2022correlation, deshpande2022association}, have examined the citation disparity between articles shared on Twitter and those that were not, finding that shared articles often receive significantly more citations (sometimes nearly double), highlighting the platform's potential impact on research visibility.

Mohammadi et al. \cite{mohammadi2018academic} surveyed 1,912 users who have tweeted journal articles and found that a significant portion of users (almost half) tweeting academic papers are non-academics, highlighting Twitter's role in spreading scholarly knowledge beyond academia and supporting its use as an indicator of public interest and cross-disciplinary impact. Knight and Kaye \cite{knight2016tweet} studied how academics and undergraduates use Twitter differently in higher education, and observed that academicians focus on reputation and public engagement, while students primarily use it to passively receive information rather than actively engage in learning. Researchers have also explored the perception of educators for using these platforms \cite{carpenter2014and}, understanding communication between instructors and students \cite{chen2012use}, its benefits for students \cite{malik2019use}, and academic library discourse on Twitter \cite{kim2012disseminates}. 
 
O'Keeffe \cite{o2019academic} highlighted that Twitter offers informal learning opportunities for academic development, but many academics still face barriers to participation. The author suggested the need to rethink academic development practices to enhance staff's digital capacity and identity. Social Networking websites support information seeking, networking, and collaboration in academics \cite{shah2017uncovering}. Despite their benefits, they remain a niche tool used by a minority of academics.

\subsection{Gender Biases in Social Media}

\textbf{Usage Behavior.} Studying the population on Twitter has become a valuable proxy for studying diverse demographic behaviors across domains, such as politics~\cite{graells2019representative, wu2023say, pena2023feminism}, sports \cite{yoon2014gender, clavio2013effects}, and scientific discourse \cite{mohammadi2018academic}. Its widespread use offers an accessible alternative to traditional survey-based methods, which are often costly, time-consuming, and difficult to scale. A survey-based study found that women tend to use social media primarily for relational purposes, such as maintaining close personal connections, while men are more likely to use these platforms to access general information \cite{krasnova2017men}. 
According to Hu and Kearney \cite{hu2021gendered}, women on social media are more likely to exhibit group-oriented behavior, such as fostering cohesion, and promote their posts rather than engaging in direct interactions. Additionally, their content often centers on family-related topics more than that of men. These observations imply that social media platforms reinforce existing disparities, even if it increases the visibility of underrepresented groups. This highlights the importance of examining how gender influences communication styles online and finding ways to strengthen online communities.

In the context of sports, Yoon et al. \cite{yoon2014gender} observed that women tend to use Twitter mainly for entertainment purposes. Meanwhile, Clavio et al. \cite{clavio2013effects} reported that female fans are more active contributors to sports team feeds than their male counterparts, particularly when it comes to sharing in-game updates, match outcomes, news about individual players, as well as engaging with contests, giveaways, and promotional ticket offers. In the political domain, Evans ~\cite{evans2016women} analyzed the Twitter activity of U.S. House representatives during the 2012 election and mid-2013, finding that female politicians discuss ``female issues" more often than their male counterparts, but their focus is not limited to these topics. In fact, women also tweet about traditionally ``male issues," including business, and in some cases, do so even more than men. Gendered patterns of communication are also evident on other platforms. On Facebook, for instance, women are more likely to share personal content (e.g., family matters), while men post more about public topics like politics and sports~\cite{wang2013gender}. Although women generally receive more feedback, posts on male-focused topics attract even more engagement, particularly when shared by women.

\textbf{Expression of Sentiments and Emotions.} Gender-based differences in the expression of emotions~\cite{hakak2017emotion} and sentiments~\cite{saxena2022introduction} offer important insights into how individuals experience community and emotional connection in online spaces. Thelwall et al.\cite{thelwall2010data} analyzed MySpace interactions and found that women both gave and received more positive comments than men, although no significant gender difference emerged for negative comments. Women expressed more positive emotions, especially in interactions with other women \cite{kivran2012joy}. A large-scale analysis of the web forum dataset found that women tend to express opinions more subjectively and are more likely than men to express both positive and negative emotions \cite{zhang2013research}. In STEM-related conversations on social media, user responses to women were generally positive, with notable spikes in sentiment during events like the International Day of Women and Girls in Science and International Women’s Day, highlighting the impact of these campaigns in shaping public perception \cite{fouad2023sentiment}.

Several studies support the notion that women are generally more emotionally expressive online, particularly with regard to positive emotions and reflecting positivity~\cite{parkins2012gender, rangel2014emotions, reychav2019emotion}. Garcia-Rudolph et al.\cite{garcia2019stroke} found that female stroke survivors on Twitter expressed significantly more positive emotions than their male counterparts, who showed higher levels of negativity. In the context of sports, both male and female fans displayed intense emotions like anger and fear, but gender differences emerged in the expression of softer emotions such as joy and sadness \cite{bagic2016sentiment}. These findings align with our study, which also observes that women tend to express stronger emotions and sentiments specifically for different events, while men’s emotional expressions are more consistent over time.

\textbf{Audience Response.} Hate speech, toxicity, and online threats have become widespread on social media platforms, prompting global concern. In response, the United Nations launched a strategy and plan of action on hate speech in 2019 to help states counter such harmful discourse while upholding freedom of expression. Recent research by Miller et al.~\cite{miller2023antisemitism} highlights the severity of this issue—following Elon Musk’s acquisition of Twitter in October 2022, antisemitic hate speech in English increased by 106\% on a weekly average, with over 325,000 tweets identified in a nine-month period.

Audience reactions on social media are not only influenced by content but also by the gender of the speaker. Studies have shown that men and women receive notably different responses online~\cite{nadim2021silencing, bartlett2014misogyny}. Ghaffari \cite{ghaffari2023discourses} studied how actress Lena Dunham presents herself on Instagram against the traditional ideas about women and analyzed 2,000 user comments. The study highlighted how women are often pressured to conform to stereotypical gender norms and suppress authentic expression to avoid online misogyny and backlash. Fuchs and Sch{\"a}fer \cite{fuchs2021normalizing} studied misogynistic and sexist hate speech targeting female politicians in Japan on Twitter, revealing that such abuse is widespread but under-researched compared to other forms of online hate. A similar pattern was observed in Spain, where the Spanish Minister of Equality received hate speech and improper comments on Twitter~\cite{iranzo2024journalists}. Pillai~\cite{pillai2022indian} interviewed women influencers to understand the gendered nature of online trolling. The interviewees mentioned that there are fewer women on social media platforms, and this underrepresentation reflects broader societal biases. They also emphasized that women are more frequently targeted when discussing sensitive topics such as feminism, gender equality, and women's rights. 
These studies collectively demonstrate how gender dynamics significantly shape audience responses online, with women often facing heightened scrutiny, harassment, and silencing, especially when engaging in public discourse or advocacy.

\section{Data Collection}
We began our study with data collection from X/Twitter\footnote{In this paper, we refer it by ``Twitter'' as the platform was called Twitter during the data collection process.}. For this purpose, we selected the top 20 computer science institutes according to the Times Higher Ranking 2022 \cite{url_timeshigher}. From the selected institutes, we manually extracted academic staff details, including their full name, academic position, official email address, and gender (as listed on the homepage or identified from pronouns or profile pictures), for various positions from \textit{Professor} to \textit{Research Associate} from university websites. Next, we manually search their Twitter handles using an exact match of full names in the Twitter ID or profile description. The academics without a direct match were excluded, resulting in a dataset of 151 female and 476 male academics.  
Tables \ref{tab:academic_univeristy_stat} and \ref{tab:academic_position_stat} present the university-level and position-level statistics of the academics present in our dataset. It can be observed that there are relatively more male academics than female academics active on Twitter, which is also a reflection of the number of female academics in Computer Science at the selected institutes. 
On average, the female-to-male academic ratio remains as \textit{1:4.3} across the top 20 institutes. Further, the trend gains momentum with the rise in academic position (refer Table \ref{tab:academic_position_stat}). 

\begin{table}[h!]
\centering
\small
\caption{Statistics of the no. of academics per university}
\label{tab:academic_univeristy_stat}
\begin{tabular}{lll}
\toprule
\textbf{University}                                & \multicolumn{1}{c}{\textbf{Male}} & \multicolumn{1}{c}{\textbf{Female}} \\ \midrule
California Institute of   Technology      & 3                         & 2                           \\
Carnegie Mellon University                & 29                        & 9                           \\ 
Columbia University                       & 17                        & 4                           \\ 
Cornell University                        & 34                        & 15                          \\ 
Georgia Institute of   Technology         & 58                        & 23                          \\ 
Harvard University                        & 13                        & 6                           \\ 
Massachusetts Institute of   Technology   & 23                        & 1                           \\ 
New York University                       & 20                        & 3                           \\ 
Princeton University                      & 28                        & 9                           \\ 
Stanford University                       & 25                        & 6                           \\
The University of Texas at   Arlington    & 14                        & 3                           \\
University of California                  & 13                        & 4                           \\
University of California San   Diego      & 18                        & 8                           \\ 
University of California,   Berkeley      & 23                        & 13                          \\ 
University of Illinois   Urbana-Champaign & 23                        & 11                          \\ 
University of Michigan                    & 31                        & 7                           \\
University of Pennsylvania                & 38                        & 6                           \\ 
University of Southern   California       & 19                        & 8                           \\ 
University of Washington                  & 36                        & 13                          \\
Yale University                           & 8                         & 0                           \\ \bottomrule
\end{tabular}
\end{table}

\begin{table}[]
\centering
\small
\caption{Statistics of the number of academics per position}
\label{tab:academic_position_stat}
\begin{tabular}{lll}
\toprule
\textbf{Academic Position}            & \textbf{Male} & \textbf{Female} \\ \midrule
Professor           & 201  & 46     \\ 
Associate Professor & 91   & 39     \\
Assistant Professor & 163  & 60     \\
Lecturer            & 4    & 1      \\ 
Adjunct Faculty     & 1    & 0      \\
Associated Faculty  & 7    & 2      \\ 
Scientist           & 5    & 3      \\ 
Research Associate  & 1    & 0      \\ \bottomrule
\end{tabular}
\end{table}

\begin{table}[]
\centering
\small
\caption{Statistics of the Twitter/X data.}\label{table:twitter-data-statistics}
\begin{tabular}{llll}
\toprule
& \textbf{Male} & \textbf{Female} & \textbf{Total} \\ \midrule
No. of Tweets & 276,002 & 63,109 & 339,111 \\ 
Avg. No. of Tweets & 585 & 420 & 502 \\ 
No. of Retweets & 118,273 & 44,324 &  162,597   \\ 
Avg. No. of Retweets & 265 & 301 & 287 \\ 
No. of Tweets with Replies & 8,462 & 2,012& 10,474 \\ 
No. of Replies & 96,284 & 13,035 &  109,319    \\ 
Avg. No. of Replies & 288 & 130 &  209     \\ 
Avg. No. of Friends & 465 & 503 & 474 \\ 
Avg. No. of Followers & 8,264 & 6,330 & 7,786 \\ \bottomrule
\end{tabular}
\end{table}

Next, we collected historical tweets for these users using Academic Twitter API, limiting the data to the latest 3500 tweets and retweets per user due to the API constraint. The data was collected in Jan 2023. 
To analyze the responses of other users to these academics, we collected all replies to the tweets and retweets posted in the year 2022. 
Table \ref{table:twitter-data-statistics} summarizes the collected data. In total, we gathered 501,708 tweets and retweets posted by the academics. While male academics posted more tweets (585 tweets per person) compared to female academics (420 tweets per person), the retweet rate remains similar regardless of gender. Further, male academics received more replies (2.2 times higher) than female academics per post on average. Additionally, female academics tend to have a lower average number of followers but a slightly higher average number of friends in comparison to male academics.


\section{Analysis and Findings}
This section presents the analysis of academics' tweets and replies to them, focusing on topics, sentiment, emotions, writing styles (using Large Language Models), and perspectives. 

\subsection{Analysis of Tweets and Retweets}

We begin by analyzing the tweets and retweets posted by academics, focusing on the topics of discussion, the sentiments and emotions expressed, and their linguistic styles.

\subsubsection{\textbf{Analyzing Topics.}}\label{topic_analysis}
\begin{figure}[]
\includegraphics[width=7cm]{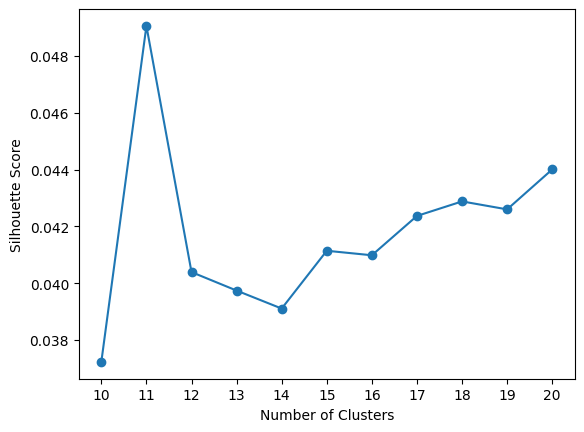}
\centering
\caption{The Silhouette scores across varying cluster size.}\label{fig:Silhouette_score} 
\end{figure}

To identify topics and the corresponding tweets or retweets, we used a clustering-based technique. We began with preprocessing the tweets by removing URLs, user mentions, emojis, and numbers using the Python library \textit{tweet-preprocessor} \cite{pypi}. 
Next, we used \textit{Twitter4SSE} \citep{di-giovanni-brambilla-2021-exploiting}, Twitter embeddings trained from the BERTweet model \citep{bertweet} for obtaining the vector representation of the text. The \textit{Twitter4SSE} converts a tweet to a 768-dimensional dense vector representing the semantic information present in the tweet. We applied the KMeans clustering \citep{Jin2010} algorithm to identify the topic clusters from the tweet embeddings. The Silhouette score (a widely used cluster evaluation metric) \citep{rousseeuw1987silhouettes} was used for unlabeled clustering to identify the optimal number of topic clusters. We randomly sampled 50,000 tweets (10\% of the tweets) and varied the number of clusters from 10-20 to find the optimal number of clusters. The optimal number of clusters was chosen based on the highest Silhouette score observed in the sample. 
Figure \ref{fig:Silhouette_score} shows the Silhouette score obtained for the sample tweets for a range of cluster sizes. It can be observed that the highest Silhouette score is obtained when the cluster size is set to 11. We used this optimal cluster size to identify the topic clusters. 

\begin{figure}[]
  \centering
  \begin{subfigure}[b]{0.47\textwidth}
    \centering
    \includegraphics[width=\textwidth]{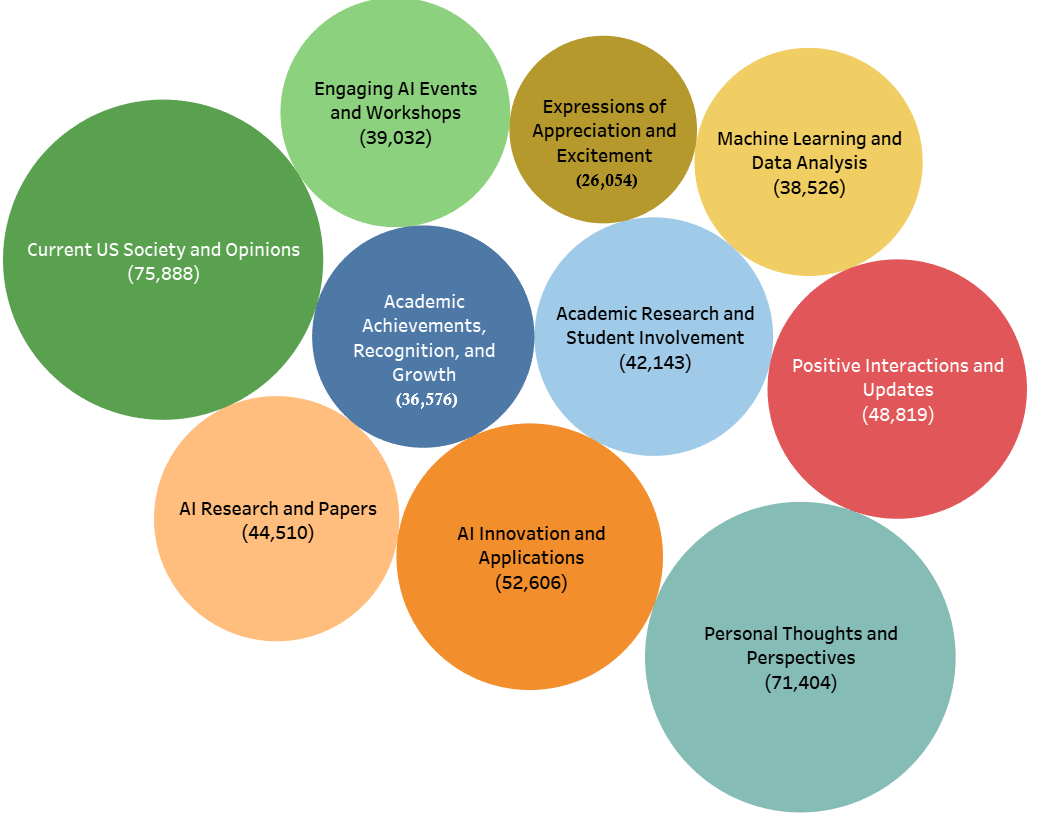}
\centering
\caption{Identified topics and number of posts per topic}\label{fig:topics}
  \end{subfigure}\\
  \vspace{3mm}
  \begin{subfigure}[b]{0.45\textwidth}
    \centering
    \includegraphics[width=\textwidth]{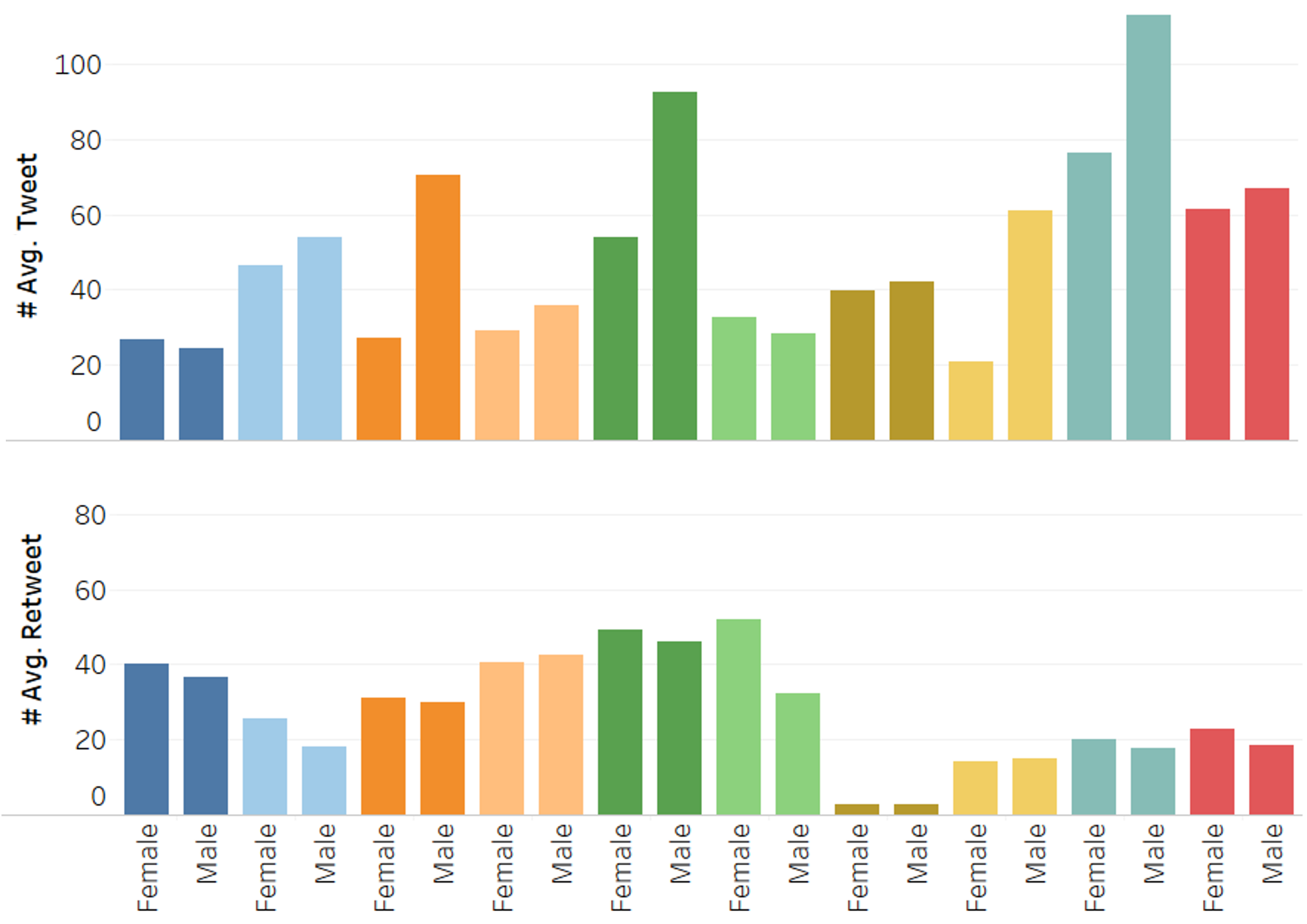}
\centering
\caption{Number of tweets and retweets posted per topic by academics}\label{fig:acadmics-topics-stat}
  \end{subfigure}\\
  \vspace{3mm}
  \begin{subfigure}[b]{0.45\textwidth}
    \centering
    \includegraphics[width=\textwidth]{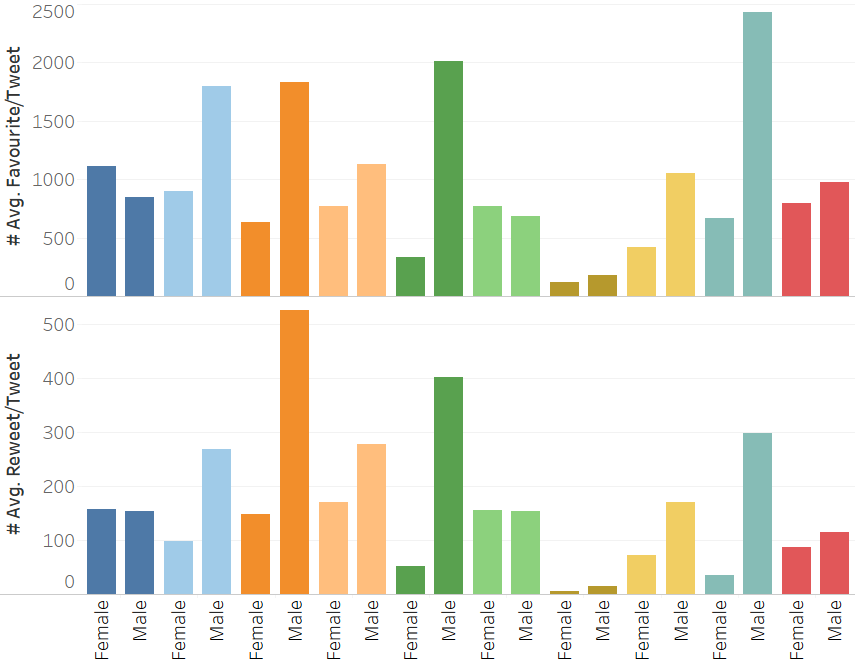}
\centering
\caption{Statistics of the reactions to the tweets posted  by the academics}\label{fig:acadmics-topics-stat-reaction}
  \end{subfigure}
  \caption{Statistics of the identified topics and tweets, retweets and response corresponding to each topic.} 
  \label{fig:topic_stats}
\end{figure}

\begin{table*}[t]
\centering
\small
\caption{Sample of topic labels for identified topics.}\label{table:topic-samples}
\begin{tabular}{p{5.8cm}p{11cm}}
\toprule
\textbf{Topic Label} & \textbf{Top 20 Words} \\ \midrule
Expressions of Appreciation and Excitement &  congrats, congratulations, thanks, thank, welcome, yes, well, deserved, awesome, nice, wow, great, good, cool, amazing, happy \\
Engaging AI Events and Workshops & talk, workshop, excited, great, amp, today, join, learning, conference, ai, work, us, looking, new, tomorrow, forward, next, thanks, week, session \\
Academic Research and Student Involvement & students, paper, one, papers, like, people, work, think, student, research, time, phd, also, get, good, many, school, know, would, faculty \\
Current USA Society and Opinions & people, trump, us, one, new, like, news, amp, would, think, covid, get, time, see, know, right, good, also, many, today \\ 
Positive Interactions and Updates & thanks, thank, great, new, see, time, today, one, good, day, work, us, happy, get, like, first, year, love, welcome, amp \\
\bottomrule
\end{tabular}
\end{table*}

To interpret each topic cluster, we extracted the top 20 frequent words (excluding the stop words \cite{nltkNLTKSearch}) from each cluster and requested Chat-GPT to identify a topic label with a maximum length of five words. We manually verified that Chat-GPT was generating meaningful topic labels using the frequent words of each topic cluster. Table \ref{table:topic-samples} provides examples of topics alongside the top 20 frequent words. Figure \ref{fig:topics} presents the 11 identified topics and the number of posts in each cluster. It can be observed that topics such as \textit{Current US Society and Opinions} and \textit{Personal Thoughts and Perspectives} are the most common, whereas the topic like \textit{Expressions of Appreciation and Excitement} forms the smallest cluster.

Figure \ref{fig:acadmics-topics-stat} shows the statistics of the topics posted by the academics. It is noticed that the retweet trend is similar for both male and female academics regardless of the topics discussed. 
However, on average, male academics tend to post the highest number of tweets on certain topics, including \textit{AI Innovation and Applications}, \textit{Current US Society and Opinions}, \textit{Machine Learning and Data Analysis}, and \textit{Personal Thoughts and Perspectives}. Only for the topic \textit{Engaging AI Events and Workshops}, the women academics tend to post slightly more tweets on average. A similar trend is observed for the retweets about the sample topic. As evident, both genders retweeted the lowest number of tweets discussing \textit{Expressions of Appreciation and Excitement} on average.

Figure \ref{fig:acadmics-topics-stat-reaction} presents the average number of retweets and favorites received by tweets posted by academics. A clear pattern emerges in how audiences react to similar topics depending on whether the tweet was posted by a male or female academic. Overall, male academics tend to receive more engagement, both in retweets and favorites, across most topics. Especially the tweets posted by male academics on topics such as \textit{Current US Society and Opinions} and \textit{Academic Research and Student Involvement} garner significantly more reactions compared to those from female academics. However, topics such as \textit{Academic Achievements, Recognition and Growth} and \textit{Engaging AI Events and Workshops} receive a slightly higher number of favorites when tweeted by female academics. Overall, the tweets posted about \textit{Expression of Appreciation and Excitement} attract the least engagement, making it the topic with the lowest average reaction across both genders.

\subsubsection{\textbf{Analyzing Sentiments.}}
Next, we analyze sentiments expressed in the tweets posted by academics. Unlike the preprocessing done for topic analysis, we only remove URLs, user mentions, and numbers and retain emojis as they explicitly convey sentiment. We used Pysentimiento \cite{perez2021pysentimiento}, a BERT architecture-based tool, to identify sentiment expressed in the tweets. This tool generates a probability distribution across three sentiment classes: \textit{positive, negative} and \textit{neutral}. To derive a single sentiment score ranging from -1 to +1, the class with the highest probability is mapped to its corresponding score, representing the overall sentiment of the tweet.

\begin{figure*}[t]
  \centering
  \begin{subfigure}[b]{\textwidth}
    \centering
    \includegraphics[width=\textwidth]{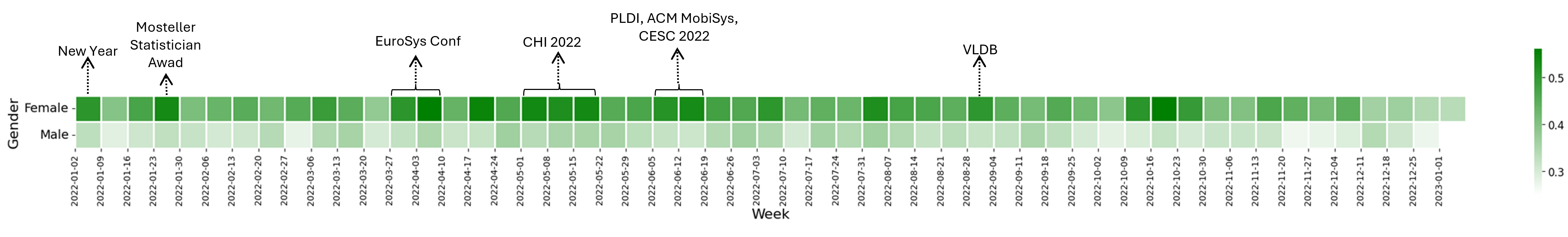}
    \caption{Average positive sentiment scores}
    \label{fig:sentiment_analysis_pos}
  \end{subfigure}
  \hfill
  \begin{subfigure}[b]{\textwidth}
    \centering
    \includegraphics[width=\textwidth]{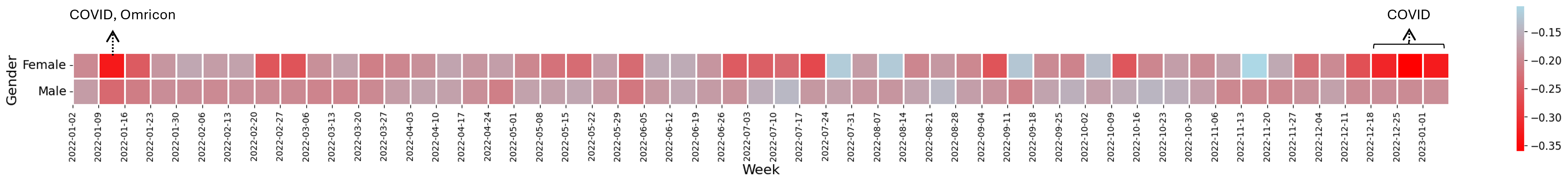}
    \caption{Average negative sentiment scores}
    \label{fig:sentiment_analysis_neg}
  \end{subfigure}
  \caption{Weekly avg. of positive and negative sentiment scores by both gender with key events identified at sentiment peaks.}
  \label{fig:sentiment_analysis}
\end{figure*}

The overall average sentiment scores indicate that female academics tend to express more positive sentiment compared to their male counterparts, while no significant difference is observed in the expression of negative sentiment. However, to gain deeper insights into the underlying reasons for these differences, we conducted a weekly sentiment analysis to examine the impact of specific events and how they influence the expression and intensity of sentiments across genders.  
To conduct time-based sentiment analysis, we calculated the weekly average of sentiment scores, excluding neutral tweets, to focus on identifying if there is a difference in the sentiment expressed by different genders. 
Figure \ref{fig:sentiment_analysis} illustrates the weekly average of positive and negative scores. Notably, women were observed posting tweets with stronger sentiments than men for any event. We manually analyzed the events associated with positive and negative sentiments for weeks with the highest sentiment scores to identify recurring events influencing these trends. For instance, women frequently shared positive tweets about general events like New Year celebrations and professional events such as conferences. Conversely, a noticeable increase in negative tweets is triggered among female academics for events like COVID-19. 

These findings are consistent with previous studies \cite{macedo2024gender, thelwall2010data, kivran2012joy} that have examined gender-based differences in Twitter usage more broadly, suggesting that the observed patterns are not unique to academics but reflect generalizable trends across user populations. As highlighted in other research, this tendency may be attributed to intrinsic characteristics in the way women express sentiments and emotions, irrespective of their professional background.

\subsubsection{\textbf{Analyzing Emotions.}}

\begin{figure}[t]
  \centering
  \begin{subfigure}[b]{0.235\textwidth}
    \centering
    \includegraphics[width=\textwidth]{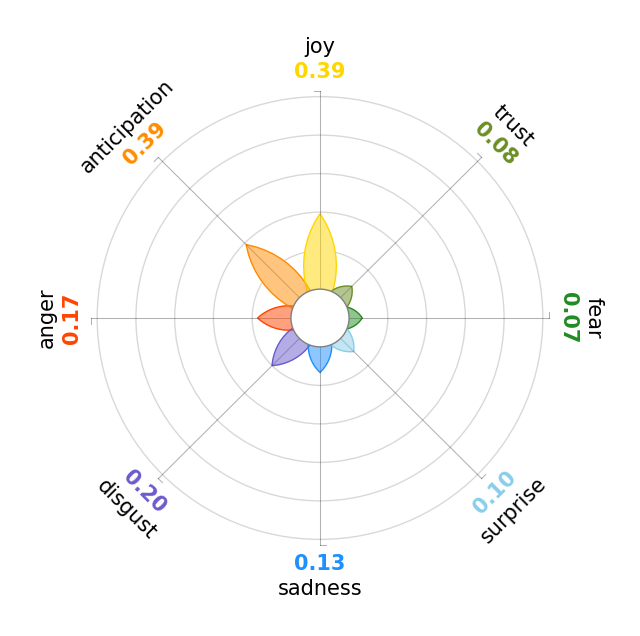}
    \caption{Men - Tweets}
    \label{fig:men_emotion_tweet}
  \end{subfigure}
  \begin{subfigure}[b]{0.235\textwidth}
    \centering
    \includegraphics[width=\textwidth]{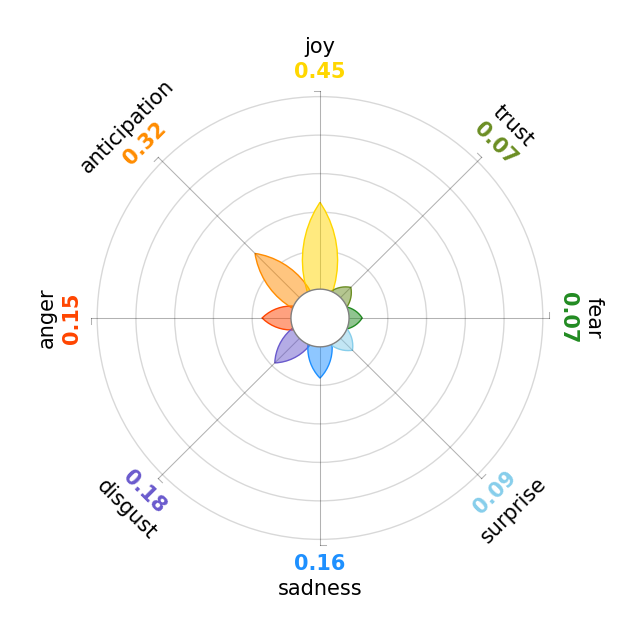}
    \caption{Women - Tweets}
    \label{fig:women_emotion_tweet}
  \end{subfigure}
  \begin{subfigure}[b]{0.235\textwidth}
    \centering
    \includegraphics[width=\textwidth]{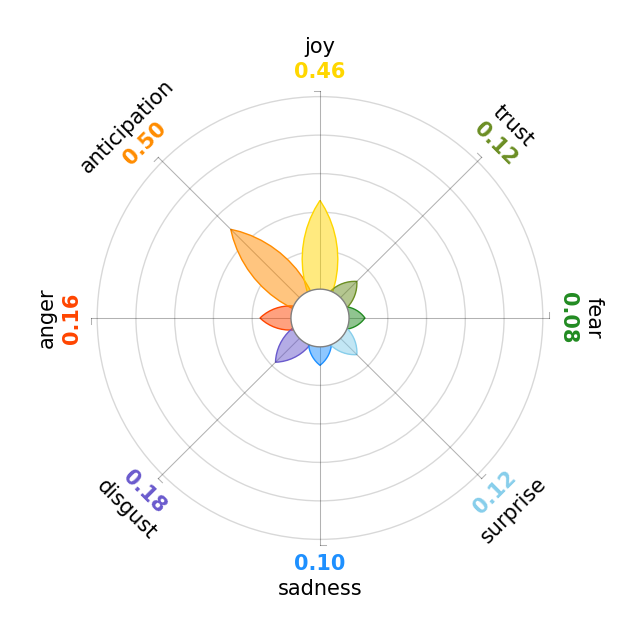}
    \caption{Men - Retweets}
    \label{fig:men_emotion_retweet}
  \end{subfigure}
  \begin{subfigure}[b]{0.235\textwidth}
    \centering
    \includegraphics[width=\textwidth]{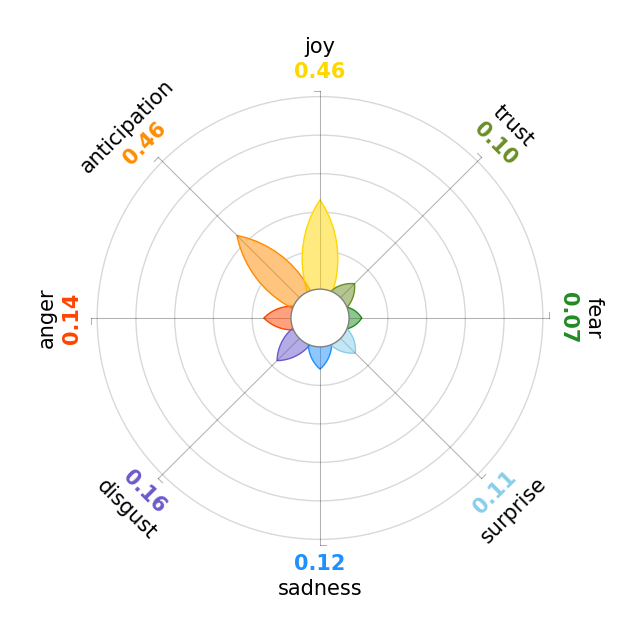}
    \caption{Women - Retweets}
    \label{fig:women_emotion_retweet}
  \end{subfigure}
  \caption{Average emotion expressed in tweets and retweets posted by male and female academics}
  \label{fig:emotion_analysis}
\end{figure}

In addition to sentiment analysis, we conduct a fine-grained emotion analysis of tweets and retweets posted by academics. For this purpose, we utilize the TweetNLP library \cite{camacho-collados-etal-2022-tweetnlp}, which classifies emotions based on the Plutchik wheel of emotions comprising eight categories: joy, trust, fear, surprise, sadness, disgust, anger, and anticipation. The emotion classifier assigns a score ranging from 0 to 1 for each emotion. 

Figure \ref{fig:emotion_analysis} illustrates the average emotion scores in tweets and retweets shared by male and female academics. Overall, joy and anticipation emerge as the most prominently expressed emotions across both genders. Notably, women exhibit higher levels of joy in their tweets, with an average score of 0.45—approximately 6\% higher than that of male academics. Conversely, male academics express anticipation more strongly, with an average score 7\% higher than their female counterparts. Beyond these differences, the distribution of other emotions remains relatively consistent between genders across both tweets and retweets. Interestingly, \textit{disgust} emotion remains the third popular emotion across both genders. Further analysis of the association between fine-grained emotions and discussion topics revealed notable patterns. The topic \textit{Current US Society and Opinions} is linked to the highest levels of negative emotions like fear, anger, disgust, and sadness. In contrast, the topic \textit{Academic Achievements, Recognition, and Growth} is most strongly associated with tweets and retweets expressing high levels of trust and joy. Meanwhile, the topic \textit{Engaging AI Events and Workshops} is predominantly connected with tweets and retweets, conveying a strong sense of anticipation.

\subsubsection{\textbf{Analyzing Writing Style.}}

We examine the writing styles of tweets posted by academics to investigate whether there are notable differences in how individuals from different genders express themselves on social media platforms. Specifically, we identify eight factors: \textit{self-praising, empathy, professional writing, discussing personal problems, politeness, sharing personal experience, stereotypical language, and expressing opinion}, which are analyzed using Large Language Models (LLMs) to capture linguistic style. The prompt \ref{prompt} was used to identify the eight writing style factors from tweets posted by the academics. The Mixtral 7x8B \cite{huggingfaceMixtral} large language model was prompted to identify the presence of each factor by responding to a \textit{Yes/No} question. Additionally, the model was allowed to respond with \textit{Not Sure} when uncertain.

\begin{prompt}[!t]
\begin{footnotesize}
\begin{verbatim}
Based on the tweet given, answer the following questions;
1. Does the author self-praise?     
2. Does the author show/express any empathy?        
3. Does the author use professional writing?  
4. Does the author discuss a personal problem?
5. Is the author being polite?     
6. Is the author sharing any personal life experience?  
7. Is the author making any stereotypical comments? 
8. Does the author express any opinion?

Respond with Yes/No/Not Sure.
\end{verbatim}
  \caption{Writing Analysis}
  \label{prompt}
\end{footnotesize}
\end{prompt}

Table \ref{tab:writing_style_analysis} shows the gender-wise percentage of \textit{Yes} response generated by the Mixtral model for each question. Notably, both male and female academics exhibit similar patterns across most factors analyzed. However, female academics are observed to be more empathetic and more likely to share personal problems and experiences compared to male academics.  

\begin{table}[]
\centering
\small
\caption{Yes responses percentage for writing style factors} 
\label{tab:writing_style_analysis}
\begin{tabular}{lll}
\toprule
\textbf{Factor}                    & \textbf{Male}  & \textbf{Female} \\ \midrule
Self-praise               & 38.7\%  & 37.8\%   \\
Empathetic                & 9.7\%   & 16.8\%   \\ 
Professional Writing      & 91.33\% & 91.8\%   \\
Discuss Personal Problem  & 6.6\%   & 11.6\%   \\
Polite                    & 93.2\%  & 94.2\%   \\
Share Personal Experience & 4.6\%   & 7.6\%    \\ 
Stereotypical             & 2.3\%   & 2.5\%    \\ 
Express Opinion           & 78.8\%  & 77.1\%   \\ \bottomrule
\end{tabular}
\end{table}

\subsection{Analysis of Replies}
In this experiment, we intend to determine whether people express their perspectives differently towards a particular gender using the replies posted to academicians' tweets. Using Google's Perspective API \cite{perspectiveapi}, we measure the intensity of five perspectives: \textit{threat}, \textit{severe toxicity}, \textit{identity attack}, \textit{insult}, and \textit{profanity}, ranging from 0 and 1.

\begin{figure*}[]
  \centering
  \begin{subfigure}[b]{0.33\textwidth}
    \centering
    \includegraphics[width=\textwidth]{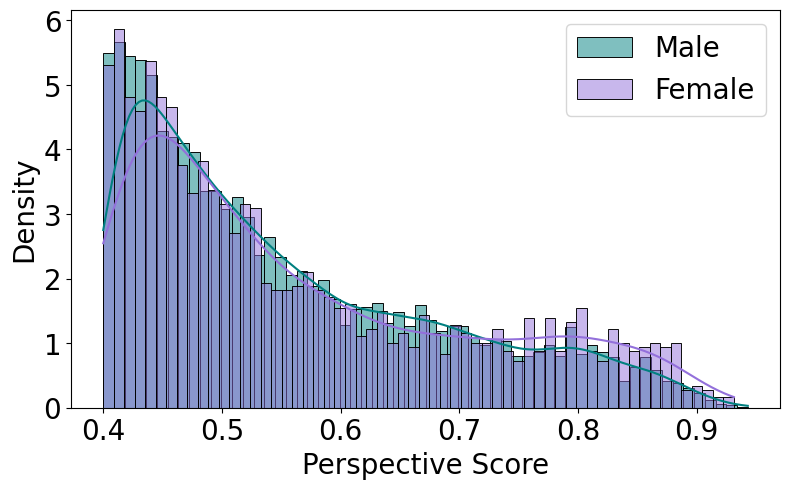}
    \caption{Threat}
    \label{fig:threat}
  \end{subfigure}
  \begin{subfigure}[b]{0.33\textwidth}
    \centering
    \includegraphics[width=\textwidth]{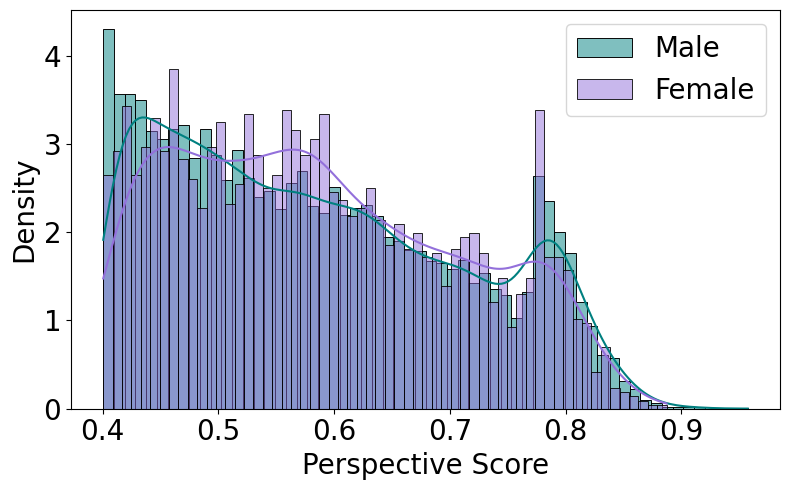}
    \caption{Severe Toxicity}
    \label{fig:severe_toxicity}
  \end{subfigure}
  \begin{subfigure}[b]{0.33\textwidth}
    \centering
    \includegraphics[width=\textwidth]{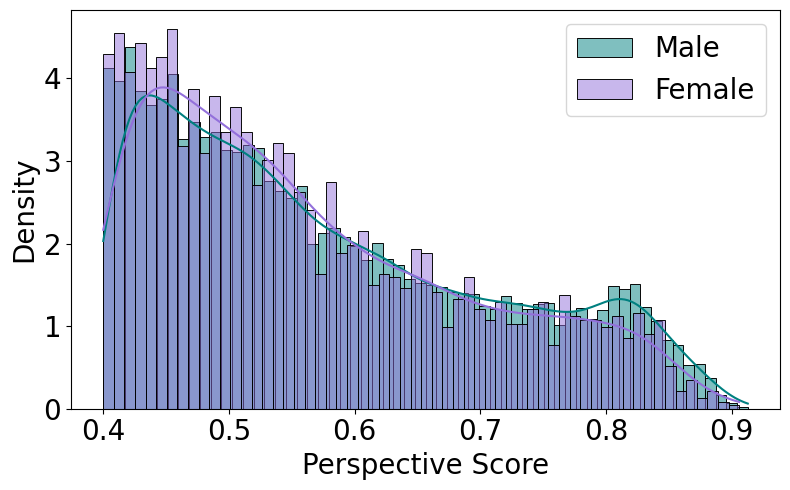}
    \caption{Identity Attack}
    \label{fig:identify attack}
  \end{subfigure}
  \caption{Density distribution of perspective scores}
  \label{fig:perspective_density distribution}
\end{figure*}

\begin{table}[]
\small
\begin{tabular}{llll}\toprule
 &                 & Male      & Female    \\ \midrule
\multirow{5}{*}{\begin{tabular}[c]{@{}l@{}}Tweet-wise Avg. \\ perspective\end{tabular}} & Threat          & $0.19 \pm 0.15$ & $0.21 \pm 0.17$ \\
& Severe Toxicity & $0.19 \pm 0.22$ & $0.19 \pm 0.23$ \\
& Identity Attack & $0.22 \pm 0.18$ & $0.23 \pm 0.19$ \\
& Insult & $0.02 \pm 0.05$ & $0.02 \pm 0.04$ \\
& Profanity &  $0.03 \pm 0.05$ & $0.03 \pm 0.05$ \\\midrule
\multirow{5}{*}{\begin{tabular}[c]{@{}l@{}}Avg. percentage \\ of replies with \\ high perspective\end{tabular}} & Threat          & 10.7\%    & 15.6\%    \\
& Severe Toxicity & 18.3\%    & 20.4\%    \\
& Identity Attack & 22.7\%    & 21.1\% \\
& Insult & 2\%    & 1\%    \\
& Profanity & 2\%    & 1\%    \\ \bottomrule
\end{tabular}
\caption{Statistics of the perspective scores of the replies}\label{tab:stat_perpective}
\end{table}

Table \ref{tab:stat_perpective} presents the tweet-wise average perspective scores and the average percentage of replies with a high perspective score (> 0.4) across five perspectives. Overall, both genders receive a low proportion of replies expressing a lower intensity of insults and profanity. However, the other three perspectives—threat, severe toxicity, and identity attack—are more frequently expressed in replies to both genders, with an average score of 0.19 - 0.23. Moreover, female academics receive a higher average percentage of replies categorized as threats and severe toxicity, while both genders experience a nearly equal percentage of identity attacks. Among these higher perspective replies, a greater number of replies with high threat scores ranging from 0.7 to 0.9 (see figure \ref{fig:threat}),  as well as more replies with severe toxicity scores between 0.5 and 0.8 (see figure \ref{fig:severe_toxicity}) compared to male academics. Conversely, male users receive a slightly higher number of replies with identity attack scores exceeding 0.8 (see figure \ref{fig:identify attack}).  
 
We conducted a regression analysis on replies with high perspective scores to further examine the language directed toward each gender and to assess whether there are statistically significant correlations beyond simple average comparisons. This approach allowed us to capture deeper patterns in linguistic behavior and determine whether gender-based differences in hostile or toxic replies persist after controlling for other factors. 
For this purpose, we train a binary classifier for each perspective to predict the gender of the original tweet's author based on the reply. Only replies with high perspective intensity (scores above 0.4) were considered. To address the imbalance in reply counts towards gender, we used undersampling by selecting an equal number of replies for the majority class (replies to male academics in our case). The data was split into a .6:.4 ratio for training and testing the classifier. For each perspective, we fine-tuned the BertTweet \cite{bertweet}, a BERT-based model, pretrained on 850M tweets. For the \textit{insult} and \textit{profanity} perspectives, there were very few replies with higher strength, which were not sufficient to train the classifier; hence, they are not considered further. 

\begin{figure}
    \centering
    \includegraphics[width=\linewidth]{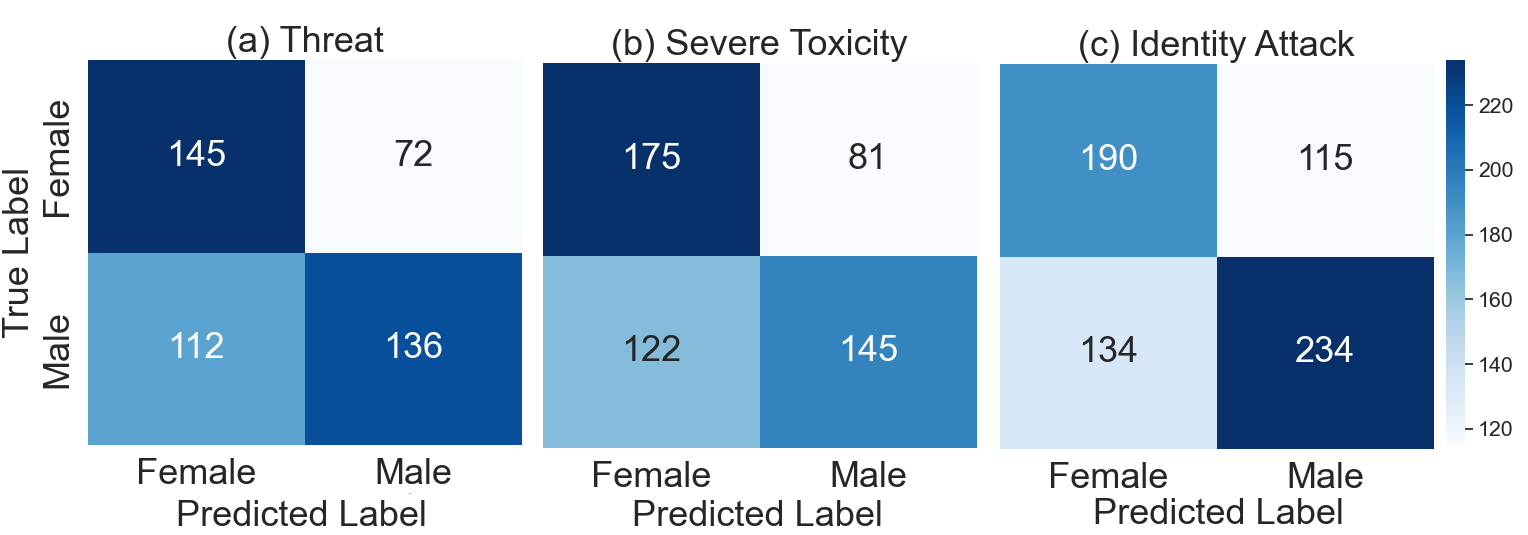}
    \caption{Confusion matrix of fine-tuned BertTweet model for Threat, Severe Toxicity, and Identity Attack perspectives.}
  \label{fig:perspective_analysis}
\end{figure}

Figure \ref{fig:perspective_analysis} presents confusion matrices of the fine-tuned model for three main perspectives: \textit{threat}, \textit{severe toxicity}, and \textit{identity attack}. Interestingly, the model identifies the threatening and severely toxic replies directed at female academics. A similar trend is observed for identity attacks targeting male academics. Results highlight that there are distinct patterns in how different perspectives are expressed toward specific genders. 

\section{Discussion and Conclusion}

This study conducts a gender-based analysis of academic communication patterns on social media platform X, especially focusing on differences in how they represent themselves and how the audience responds to them. Topic-based analysis reveals notable differences, such as male academics being more active in tweeting about societal topics, subject matter areas such as artificial intelligence and machine learning, and personal thoughts. Despite similar retweet patterns, male-authored tweets tended to receive significantly more engagement, especially on topics like Current US Society and Opinions, whereas female-authored tweets drew slightly more favorites on topics tied to Academic Achievements. Understanding topic-based influence dynamics will enable more effective identification of influential users across different contexts \cite{panchendrarajan2023topic}.

The sentiment analysis of tweets posted by academics reveals that female academics consistently express stronger sentiments—both positive and negative—than their male counterparts across various events. The analysis shows that women tend to post more positively during occasions like New Year celebrations and professional events such as conferences. In contrast, spikes in negative sentiment, particularly among female academics, were associated with impactful events such as the COVID-19 pandemic. These patterns suggest that women academics engage more emotionally with both personal and professional experiences shared on Twitter. 

In addition to sentiment analysis, the study conducted a fine-grained emotion analysis of tweets and retweets posted by academics using the TweetNLP library. Results show that joy and anticipation are the most expressed emotions across both genders, with women exhibiting notably higher joy and men showing stronger anticipation. While other emotions such as fear, anger, and sadness are similarly distributed, disgust surprisingly emerges as the third most prominent emotion for both genders. Further linking emotions to discussion topics reveals clear emotional patterns: tweets about Current US Society and Opinions contain the highest levels of negative emotions, while Academic Achievements are rich in trust and joy, and topics like Engaging AI Events and Workshops generate strong anticipation. Our results are in correlation with previous studies \cite{parkins2012gender, rangel2014emotions, reychav2019emotion, bagic2016sentiment, macedo2024gender} highlighting that irrespective of topics or user types (academics/non-academics), females show strong sentiments, higher level of soft emotions like joy, and more positivity on online social media platforms. 

Further, the writing style analysis using large language models indicates that female academics are more likely to show empathy and share personal problems in their posts. There was no significant difference in other factors compared. Individuals from both genders engage in self-praising behavior on social media platforms. However, a more fine-grained, tweet-level content analysis could help uncover whether there are significant gender-based differences in the topics or forms of self-praise. Previous research has suggested that men tend to self-promote or share achievements more frequently than women \cite{peng2022gender}. In contrast, our findings reveal no significant gender difference in self-praising behavior, which may be attributed to the homogeneity of our datasets as all users are academics from the same professional field, with similar educational backgrounds and comparable academic profiles.

To understand audience response to academics, we analyze whether replies to tweets by male and female academics differ in the expression of harmful language, focusing on five perspectives: threat, severe toxicity, identity attack, insult, and profanity. Our findings show that while insults and profanity are generally low in intensity for both genders, replies to female academics more frequently contain higher levels of threats and severe toxicity. Identity attacks occur at similar rates for both genders but with a slight skew toward more intense attacks directed at male academics. We also perform regression analysis using a fine-tuned BertTweet model on replies to study the distinction in linguistic patterns based on the gender of the tweet's author. The model reliably identifies threatening and toxic replies targeting women and identity attacks toward men, suggesting that gender plays a significant role in the nature of online hostility experienced by academics. These findings underscore gendered disparities in how harmful discourse manifests in academic Twitter conversations.
These findings shed light on how gender dynamics shape online communication among academics and the audience's interactions with them on social media, emphasizing the need for fostering a more inclusive and supportive digital environment for scholarly discourse. 

We recognize several limitations in our study. First, the collected dataset is a small subset of academics on Twitter and might not fully represent the whole set. Additionally, it might be skewed toward specific demographics—primarily white, male, and western populations, limiting the generalizability of our findings. However, as we mentioned earlier, due to having a fair comparison from a gender perspective, we limit the academicians belonging to the selected set of universities. Moreover, the gender of a person is identified using name, profile photo, and pronouns on the homepage, if provided. The study is limited to two genders, and in the future, one should expand to get a broader perspective. Still, platforms like Twitter provide key insights for promoting inclusivity through improved moderation, reporting, design, and support for those facing online harassment.
Second, while sentiment analysis via APIs provides a scalable method for analyzing emotions, it has some limitations. Previous works showed that occasionally misclassified sarcastic or ironic tweets as positive \cite{gonzalez2011identifying, saxena2022recent}. To address this, we employed a multi-layered approach that included manual validation, improving the reliability of our emotion classification. Lastly, the rapidly changing nature of social media introduces further challenges, such as evolving platform policies, shifting user behavior, and emergent trends, all of which may affect the consistency and reproducibility of results over time.

In future work, we aim to extend our study to a more granular level by analyzing the propagation dynamics of posts authored by different genders. While the current study focuses primarily on the nature of replies directed toward male and female users, our upcoming research will investigate how individuals of different genders respond to one another—for example, whether female users show greater empathy in their interactions with other females. We also plan to examine the presence of homophily in these discussions, exploring whether users are more likely to engage with those of the same gender. Additionally, given the increasingly limited access to Twitter data, we will consider alternative platforms such as Bluesky, which offers free and open access to user data, for future analysis.

\section*{Acknowledgement}
We would like to thank Aditya Ashok Kumar Padhy for discussing and providing valuable insights in sentiment and perspective analysis.

\bibliographystyle{ACM-Reference-Format}
\bibliography{mybib.bib}

\end{document}